\documentclass[11pt,preprintnumbers,aps,amssymb,nofootinbib,amsmath]
{revtex4}
\usepackage{epsfig,epsf}
\usepackage{graphicx}
\usepackage{verbatim}
\usepackage{epstopdf}
\usepackage{bm} % puts greek and math symbols in boldface using \bm
\newcommand{\beq}{\begin{equation}}
\newcommand{\eeq}{\end{equation}}
\newcommand{\bea}{\begin{eqnarray}}
\newcommand{\eea}{\end{eqnarray}}

\def\eq#1{{(\ref{#1})}}
\def\fig#1{{Fig.~\ref{#1}}}

\def\sec#1{{Sec.~\ref{#1}}}

\newcommand{\as}{\alpha_s}
\newcommand{\bas}{\bar\alpha_s}

\def\b#1{\mathbf{#1}}
\newcommand{\un}{\underline}
\begin{document}

\preprint{RBRC-729}

%~\vspace{1cm}

\title{Gluon multiplicity in coherent diffraction of onium on a heavy nucleus.}

\author{Yang Li$\,^a$ and Kirill Tuchin$\,^{a,b}$\\}

\affiliation{
$^a\,$Department of Physics and Astronomy, Iowa State University, Ames, IA 50011\\
$^b\,$RIKEN BNL Research Center, Upton, NY 11973-5000\\}

\date{\today}


\begin{abstract}
We derive the cross section for the diffractive gluon production in high energy onium-nucleus collisions that includes the low-$x$ evolution effects in the rapidity interval between the onium and the produced gluon and in the rapidity interval between the gluon and the target nucleus. We analyze our result in two limiting cases: when the onium size is much smaller  than the saturation scale and when its size is much larger  than the saturation scale. In the later case the gluon multiplicity is very small in the quasi-classical case and increases when the low-$x$ evolution effects in onium become significant. We discuss the implications of our result for the RHIC, LHC and EIC phenomenology.

\end{abstract}

\maketitle

%%%%%%%%%%%%%%%%%%%%%%%%%%%%%%%%%%%%%%%%
\section{Introduction}\label{sec:intr}

Diffractive gluon production in high energy pA collisions and in  Deep Inelastic Scattering (DIS) is  a sensitive probe of the Color Glass Condensate \cite{McLerran:1993ni,Jalilian-Marian:1997jx,Jalilian-Marian:1997gr,Jalilian-Marian:1998cb,Jalilian-Marian:1997dw,Kovner:2000pt,Iancu:2000hn,Iancu:2001ad,Iancu:2001md,Ferreiro:2001qy}  characterized by high parton density and gluon saturation \cite{Gribov:1983tu,Mueller:1986wy}.
Diffractive processes played a pivotal role in identifying the first signatures of the gluon saturation in DIS at HERA \cite{GolecBiernat:1998js,GolecBiernat:1999qd,Gotsman:1999vt,Gotsman:2000gb,Gotsman:2002zi,Levin:2002fj,Levin:2001pr,Kovchegov:1999kx,Bartels:2002cj}. They are of great interest as a tool for studying the low-$x$ dynamics in pA collisions at RHIC and LHC as well as  in DIS at the proposed EIC. 
Study of high parton densities in deuteron -- gold collisions at RHIC has provided many novel insights into the structure of nuclear matter and has been focused on inclusive processes. By triggering on hadrons in the deuteron fragmentation region (``forward'' rapidity) one is able to access very low values of Bjorken $x$ that are sensitive to the gluon saturation. 
Investigation of energy, rapidity, centrality, and transverse momentum dependence of various production channels offers an opportunity to attain a better understanding of the nuclear and hadron structure at low $x$. Among the channels which have been discussed in this context are total hadron multiplicities \cite{Kharzeev:2000ph,Kharzeev:2001gp,Kharzeev:2001yq,Kharzeev:2002ei}, incusive production of gluons  \cite{Kovchegov:1998bi,Kovchegov:2001sc,Braun:2000bh,Dumitru:2001ux,Blaizot:2004wu,Kharzeev:2002pc,Kharzeev:2003wz,Kharzeev:2004yx,Baier:2003hr,Iancu:2004bx}, heavy quarks \cite{Kharzeev:2003sk,Gelis:2003vh,Tuchin:2004rb,Blaizot:2004wv,Kovchegov:2006qn,Tuchin:2007pf,Kharzeev:2005zr}, valence quarks  \cite{Gelis:2001da,Dumitru:2002qt}, prompt photons \cite{Jalilian-Marian:2005zw}, di-leptons  \cite{Baier:2004tj,Jalilian-Marian:2004er,Betemps:2004xr} and identified hadrons \cite{Li:2007zzc}  (for a review see e.g.\ \cite{Jalilian-Marian:2005jf,Tuchin:2006hz}).  Diffractive production in pA collisions offers another avenue for exploring the low-$x$ dynamics. Motivated 
by a possibility to measure the diffractive production in pA collisions at RHIC and LHC we analyze in this paper diffractive gluon production in onium--heavy nucleus collisions. We intentionally avoid discussing the ``dipole content" of the proton light-cone ``wave function" and concentrate entirely on quantities that can be  calculated in perturbation theory.  Our results can be equally well applied to  
diffractive gluon production in DIS in which the light-cone ``wave function" of the virtual photon is well-known.  Diffractive gluon production in DIS
 has been discussed in many publications \cite{Wusthoff:1997fz,GolecBiernat:1999qd,Bartels:1999tn,Kopeliovich:1999am,Kovchegov:2001ni,Munier:2003zb,Marquet:2004xa,GolecBiernat:2005fe,Marquet:2007nf} and has been limited to the quasi-classical approximation and/or phenomenological models. In this paper we go beyond the quasi-classical approximation and include the low-$x$ evolution effects at all rapidity intervals.

The paper is structured as follows. In Sec.~\ref{sec:model}  we review  the result for   diffractive gluon production in the quasi-classical approximation derived in \cite{Kovchegov:2001ni,Kovner:2001vi,Kovner:2006ge}. In \sec{sec:evolu} we  generalize these results by including the effect of quantum evolution. We consider separately the case when the rapidity gap $Y_0$ between the produced gluon and the target  equals the gluon's rapidity $y$ (\fig{fig:diffract1}) and a more general case when $y\ge Y_0$ (\fig{fig:diffract2}). The corresponding cross sections are given by  \eq{main2} and \eq{main3}  in terms of the dipole distribution in proton $n_p(\b r, \b r',\b b, y)$, the forward dipole-nucleus scattering amplitude $N(\b r, \b b, y)$ and diffractive dipole-nucleus scattering amplitude  $N_D(\b r, \b b, y;Y_0)$. In Sec.~\ref{sec:dipole-ev} we review the main properties of $n_p(\b r, \b r',\b b, y)$ and $N(\b r, \b b, y)$ in the linear regime and in the saturation regime and demarcate the kinematic landscape. We then proceed in \sec{sec:section} by performing analysis of the diffractive gluon production in the quasi-classical approximation in various kinematic regions. The results are displayed  in \eq{va5} and \eq{vb5}. In \sec{sec:lowx} we do similar analysis in the case of low-$x$ evolution, see  \eq{via6} and \eq{vib3}.  We summarize and discuss the phenomenological importance of  the obtained results in \sec{sec:summary}.

%%%%%%%%%%%%%%%%%%%%%%%%%%%%%%%%%%%%%%%%
\section{Diffractive gluon production in the quasi-classical approximation}\label{sec:model}

%%%%%%%%%%%%%%%
%\subsection{Quasi-classical approximation}

%%%%
\begin{figure}[ht]
  \includegraphics[width=11cm]{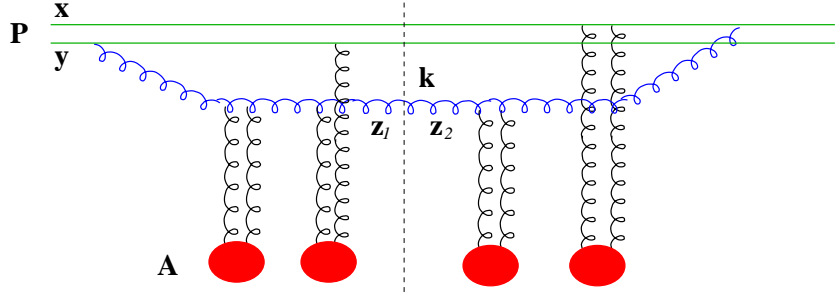}
\caption{One of the diagrams contributing to the diffractive gluon production in onium (P) --  heavy nucleus (A) collisions in the quasi-classical approximation. Notations are explained in the text.}
\label{fig:dif-class}
\end{figure}
%%%%%

The quasi-classical approximation of the hadron-nucleus interactions is valid when a typical parton coherence length $l_c$ is much larger than the nuclear size $R_A$ in the nucleus rest frame. The former is approximately given by $l_c\approx \frac{1}{2m_Nx}$, where $m_N$ is a nucleon mass. It follows that the quasi-classical approximation holds for $x\lesssim \frac{1}{R_Am_N}$. Owing to  the large coherence length, the process of diffractive production can be considered as proceeding in two stages: gluon emission long time before the collision followed by the instantaneous interaction. This picture is particularly simple in the transverse configuration space since  the parton transverse coordinates do not change in the course of instantaneous interaction. As the result, the cross section can be represented as a convolution of the proton's light-cone ``wave-function" and the scattering amplitude in the transverse configuration space, see \fig{fig:dif-class}. In the quasi-classical approximation, the cross section for the diffractive gluon production in onium--heavy nucleus collisions has been derived in \cite{Kovchegov:2001ni,Kovner:2001vi,Kovner:2006ge}. Using notations of \cite{Kovchegov:2001ni}, see \fig{fig:dif-class}, it reads:
\bea\label{xsectQ}
\frac{d\sigma(k,y)}{d^2k\, dy}&=&\frac{\as
C_F}{\pi^2}\frac{1}{(2\pi)^2}\,\int d^2b\, d^2z_1\,
d^2z_2\,\left(\frac{\b z_1-\b x}{|\b z_1-\b x|^2}-
 \frac{\b z_1-\b y}{|\b z_1-\b y|^2}\right)\cdot
 \left(\frac{\b z_2-\b x}{|\b z_2-\b x|^2}-
 \frac{\b z_2-\b y}{|\b z_2-\b y|^2}\right)\nonumber\\
 &&\times\,e^{-i\b k\cdot(\b z_1-\b z_2)}\,
 \left( e^{-P(\b x, \b y, \b z_1)}-e^{-\frac{C_F}{4N_c}(\b x-\b y)^2Q_{s0}^2}\right)
  \left( e^{-P(\b x, \b y, \b z_2)}-e^{-\frac{C_F}{4N_c}(\b x-\b y)^2Q_{s0}^2}\right)
  \,,
\eea
where $\b x$ and $\b y$ are the transverse coordinates of quark and antiquark, $\b z_1$, $\b z_2$  are the transverse coordinates of the gluon in the amplitude and the complex-conjugate amplitude correspondingly, see \fig{fig:dif-class}. The $q\bar qg$ propagator reads \cite{Kopeliovich:1998nw,Kovner:2001vi,Tuchin:2004rb}
\beq\label{P}
\exp\{-P(\b x,\b y,\b z)\}=\exp\left( -\frac{1}{8}(\b
x-\b z)^2Q_{s0}^2 -\frac{1}{8}(\b y-\b
z)^2Q_{s0}^2+\frac{1}{8N_c^2}(\b x-\b y)^2Q_{s0}^2\right)\,.
\eeq
The \emph{gluon} saturation scale is given by
\beq\label{Qsat}
Q_{s0}^2=\frac{4\pi^2\as N_c}{N_c^2-1}\,\rho\, T(\b b)\, xG(x,1/\b
r^2)\,,
\eeq
where $\rho$ is the nuclear density, $T(\b b)$ is the nuclear thickness function as a function of the impact parameter $\b b$. The gluon distribution function reads
\beq\label{xG}
xG(x,1/\b
r^2)=\frac{\as C_F}{\pi}\ln \frac{1}{\b r^2 \Lambda^2}\,,
\eeq
with
$\Lambda$ being some non-perturbative momentum scale characterizing the nucleon's wave function.

In the framework of the dipole model \cite{dip}, the gluon evolution is easily accounted for in the large $N_c$ approximation. Let us introduce the forward elastic dipole--nucleus scattering amplitude $N(\b r,\b b,Y)$. In the quasi-classical approximation it reads \cite{Mue}
\beq\label{NQ}
N(\b r,\b b,0)= 1-e^{ -\frac{1}{8}\b r^2\,
Q_{s0}^2}\,.
\eeq
At large $N_c$, \eq{xsectQ}, \eq{P} and \eq{NQ} yield
\bea\label{xsectQ2}
&&\frac{d\sigma^{pA}(k,y)}{d^2k\, dy}=\frac{\as C_F}{\pi^2}\frac{1}{(2\pi)^2}\,\int d^2b\, d^2z_1\, d^2z_2\,e^{-i\b k\cdot(\b z_1-\b z_2)}\,\nonumber\\
&&
\times \left(\frac{\b z_1-\b x}{|\b z_1-\b x|^2}-
 \frac{\b z_1-\b y}{|\b z_1-\b y|^2}\right)\cdot
 \left(\frac{\b z_2-\b x}{|\b z_2-\b x|^2}-
 \frac{\b z_2-\b y}{|\b z_2-\b y|^2}\right)\,
 \nonumber\\
 &&
 \times \left[ N(\b x-\b y,\b b,0)-N(\b x-\b z_1,\b b,0)-N(\b y-\b z_1,\b b,0)+N(\b x-\b z_1,\b b,0)N(\b y-\b z_1,\b b,0)\right]\nonumber\\
  &&
\times \left[ N(\b x-\b y,\b b,0)-N(\b x-\b z_2,\b b,0)-N(\b y-\b z_2,\b b,0)+N(\b x-\b z_2,\b b,0)N(\b y-\b z_2,\b b,0)\right]
  \,.
\eea

Integrating \eq{xsectQ2} over all transverse momenta  yields delta function $(2\pi)^2\delta(\b z_1-\b z_2)$. Hence, the total cross section per unit rapidity
reads after a simple calculation 
\bea\label{sectot}
&&\frac{d\sigma(y)}{ dy}=\frac{\as C_F}{\pi^2}\,\int d^2b\, d^2z\, \frac{(\b x-\b y)^2}{(\b x-\b z)^2(\b y-\b z)^2}\nonumber\\
 &&
 \times \left[ N(\b x-\b y,\b b,0)-N(\b x-\b z,\b b,0)-N(\b y-\b z,\b b,0)+N(\b x-\b z,\b b,0)N(\b y-\b z,\b b,0)\right]^2\,.\nonumber
\eea

%%%%%%%%%%%%%%%
\section{Including quantum evolution}\label{sec:evolu}

When the collision energy becomes high enough, multiple gluon emission becomes possible. Parametrically, each gluon emission brings in a factor $\as\ln(1/x)$. Accordingly, quantum evolution takes place when $x\lesssim e^{-\frac{1}{\as}}$.  Let the incident onium be characterized by the two-vector $\b r$. In course of evolution dipoles of different sizes are produced until eventually a dipole of size $\b r'$ emits a gluon at rapidity $y$ with transverse momentum $\b k$. 
In terms of the Regge theory, evolution in the rapidity interval between the original onium and the emitted gluon corresponds to exchange of a single Pomeron, in agreement with the AGK cutting rules \cite{Abramovsky:1973fm,Kovchegov:2001sc}. 
Afterwards, i.e.\ in the rapidity interval between the emitted gluon and the target nucleus, evolution is non-linear and corresponds to exchange of diffractively cut fan diagram, see \fig{fig:diffract1} and \fig{fig:diffract2}. The general method for including the non-linear low-$x$ evolution into inclusive processes in the dipole model framework was derived in \cite{Kovchegov:2001sc} and is applied later in this section.  

We would like to separately consider the following two cases: (i) rapidity gap $Y_0$ equals the produced gluon rapidity $y$, see \fig{fig:diffract1}, and (ii) a more general case $Y_0\le y$, see \fig{fig:diffract2}. In later sections we will focus our attention on the former case. 
 The reason is that experimentally, diffractive production is usually measured per unit of invariant mass of the diffractively produced system (rather than per $dk^2$). The invariant mass is given by $M^2=k^2/x$ where $\b k$ and $x=e^{-(Y-y)}$ refer to the slowest particle in the gluon cascade originating from proton. That being the case, it is sufficient to consider production of a gluon adjacent to the rapidity gap as depicted in \fig{fig:diffract1}. This case corresponds to the rapidity gap $Y_0$ being equal the rapidity of the produced gluon $Y_0=y$. 

%%%%%
\subsection{Gluon production with $y=Y_0$}

%%%%
\begin{figure}[ht]
      \includegraphics[height=8cm]{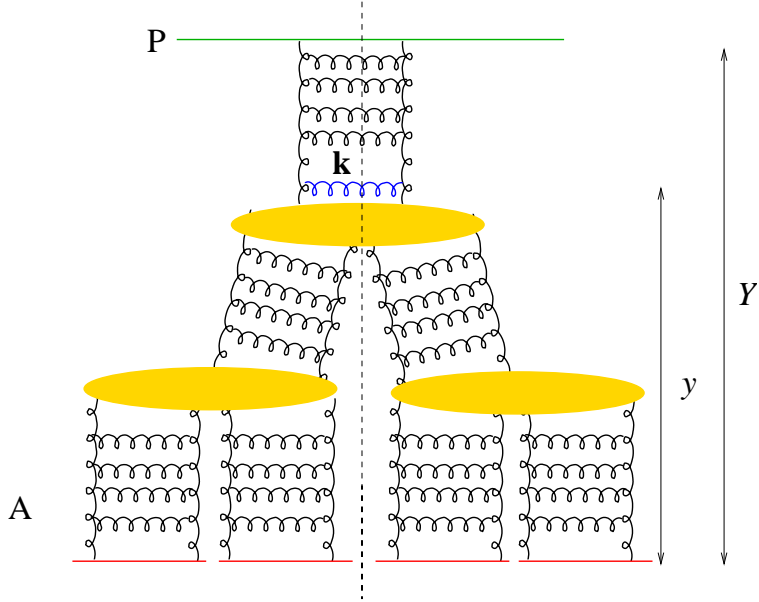}
\caption{Fan diagram describing the diffractive gluon production in onium (P) -- heavy nucleus (A) collisions with the rapidity gap being equal to the rapidity of the produced gluon.}
\label{fig:diffract1}
\end{figure}
%%%%%
The relevant fan diagram is displayed in \fig{fig:diffract1}. We include the evolution effects using the method derived in \cite{Kovchegov:2001sc}.  We obtain the following generalization of \eq{xsectQ2}:
\beq\label{main}
\frac{d\sigma(k,y)}{d^2k\,
dy}=\frac{\as C_F}{\pi^2}\frac{1}{(2\pi)^2}\,\int d^2b\,d^2B\, \int
d^2r'\,  n_1(\b r,\b r',\b B-\b b,Y-y)\,  |\b I(\b r',\b k,y)|^2\,,
\eeq
where
\bea\label{Igen}
&&\b I(\b x'-\b y',\b k,y)=\int d^2z_1\, e^{-i\b k\cdot \b z_1} \left(\frac{\b z_1-\b x'}{|\b z_1-\b x'|^2}- \frac{\b z_1-\b y'}{|\b z_1-\b y'|^2}\right)\nonumber\\
&&\times\left[ N(\b x'-\b y',\b b,y)-N(\b x'-\b z_1,\b b,y)-N(\b
y'-\b z_1,\b b,y)+N(\b x'-\b z_1,\b b,y)N(\b y'-\b z_1,\b
b,y)\right]\,.
\eea
Here $n_1(\b x-\b y,\b x'-\b y',\b B-\b b,Y-y)$ has the meaning of the number of dipoles of size $\b x'-\b y'$ at rapidity $Y-y$ and impact parameter $\b b$ generated by evolution from the original dipole $\b x-\b y$ having rapidity $Y$ and impact parameter $\b B$ \cite{dip}. It satisfies the BFKL equation \cite{Kuraev:1977fs,Balitsky:1978ic} 
\bea\label{BFKL}
 &&\frac{\partial n_1(\b x-\b y,\b x'-\b y',\b b, y)}{\partial y}=\frac{\as N_c}{2\pi^2}\int d^2z \, \frac{(\b x-\b y)^2}{(\b x-\b z)^2(\b y-\b z)^2}\nonumber\\
&&
\big[
n_1(\b x-\b z,\b x'-\b y',\b b,y)+n_1(\b y-\b z,\b x'-\b y',\b b,y)-n_1(\b x-\b y,\b x'-\b y',\b b,y)\big]\,,
\eea
with the initial condition
\beq\label{in.cond}
n_1(\b r,\b r',\b b,0)=\delta(\b r-\b r')\,\delta(\b b)\,,
\eeq
where we denoted $\b r=\b x-\b y$ and $\b r'=\b x'-\b y'$.

The forward elastic dipole--nucleus scattering amplitude satisfies the nonlinear BK equation \cite{Balitsky:1995ub,Kovchegov:1999yj}
 \bea\label{BK}
 &&\frac{\partial N(\b x-\b y,\b b, y)}{\partial y}=\frac{\as N_c}{2\pi^2}\int d^2z \, \frac{(\b x-\b y)^2}{(\b x-\b z)^2(\b y-\b z)^2}
 \big[
N(\b x-\b z,\b b,y)\nonumber\\
&&
+N(\b y-\b z,\b b,y)-N(\b x-\b y,\b b,y)-N(\b x-\b z,\b b,y)N(\b y-\b z,\b b,y)
\big]\,,
\eea
with the initial condition given by \eq{NQ}. In writing both equations \eq{BFKL}
and \eq{BK} we assumed that the absolute value of impact parameter $\b b$ is much larger than the typical dipole size. This is a justified approximation for a scattering off a heavy nucleus.

In order to keep expressions as compact as possible, it is convenient to assume that the nuclear profile is cylindrical. This simple model allows correct identification of the atomic number (i.e.\ centrality) dependence of the cross sections. An explicit impact parameter dependence, which is required for numerical analysis, can be easily restored in the final expressions. Since we are not concerned here with the details of the impact parameter dependence, we integrate \eq{BFKL} over $\un b$. The quantity
\beq\label{defnp}
n_p(\b r,\b r',y)=\int d^2b\,  n_p(\b r,\b r',\b b, y)
\eeq
in turn satisfies the BFKL equation with the initial condition
\beq\label{defnpic}
n_p(\b r,\b r',0)=\delta(\b r-\b r')\,.
\eeq
In terms of $n_p(\b r, \b r', y)$, \eq{main} reads
\beq\label{main2}
\frac{d\sigma(k,y)}{d^2kdy} = \frac{\as C_F}{\pi^2}\frac{1}{(2\pi)^2}\,S_A\int d^2r' \, n_p(\b r, \b r', Y-y)\, |\b I(\b r',\b k,y)|^2\,,
\eeq
where $S_A$ is the cross sectional area of the interaction region. 

The total cross section for diffractive gluon production is convenient to write in the following form 
\bea\label{main-tot}
\frac{d\sigma(y)}{dy} &=& \frac{\as C_F}{\pi^2}S_A\int d^2\b r' \, n_p(\b r, \b r', Y-y)\, \int d^2w\, \frac{\b r'^2}{(\b w-\b r')^2\,\b w^2}\nonumber\\
&& \times
\left[ N(\b r',\b b,y)-N(\b w-\b r',\b b,y)-N(\b w,\b b,y)+N(\b w-\b r',\b b,y)N(\b w,\b b,y)\right]^2\,.
\eea
where we introduced a new variable $\b w=\b z-\b y'$ such that $w$ is the size of one of the daughter dipoles formed by emission of a gluon at point $\b z$ by a parent dipole $\b r'=\b x'-\b y'$.

%%%%%%%%%%%%%%%%%%%%%%%%%%%%%%
\subsection{Diffractive production with  $Y_0<y$}\label{sec:bezgovna}

%%%%
\begin{figure}[ht]
      \includegraphics[height=8cm]{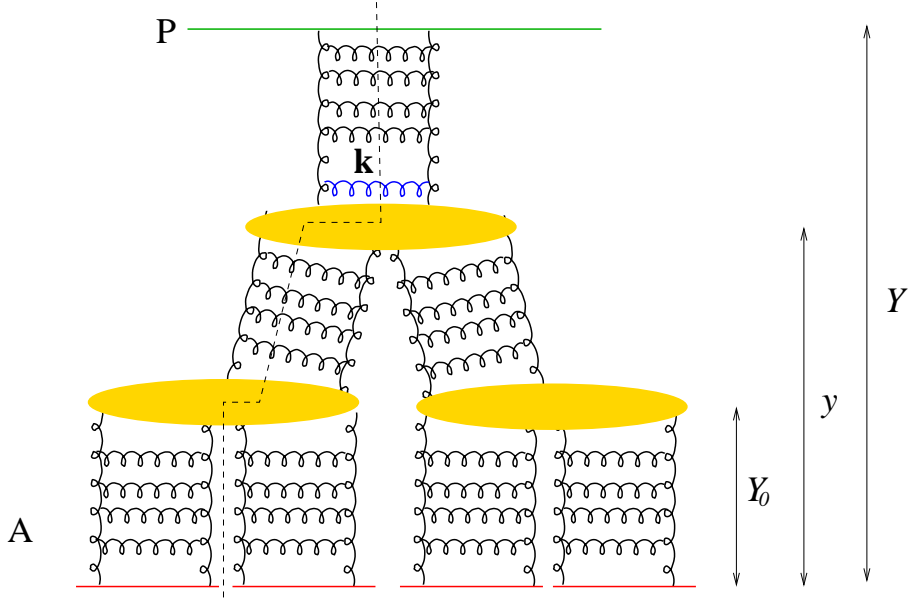}
\caption{Fan diagram for the diffractive gluon production in onium (P) -- heavy nucleus (A) collisions with rapidity gap $Y_0$ smaller than the gluon rapidity $y$.}
\label{fig:diffract2}
\end{figure}
%%%%%

So far we have been concentrating on a case in which the rapidity of the produced gluon $y$ coincides with the rapidity gap $Y_0$ in a diffractive event. In this case the diffractive scattering amplitude $N_D(\b r, \b b, y, Y_0)$ coincides with the square of the forward elastic scattering amplitude $N(\b r, \b b, y)$. In principle, a question may arise about the diffractive production of a gluon with $y> Y_0$. Such process is  shown in \fig{fig:diffract2}. For the processes in which the transverse coordinate of the gluon in the amplitude $\b z_1$ is approximately the same as  its coordinate in the c.c\ amplitude $\b z_2$, the corresponding cross section is still given by \eq{main2}
\beq\label{main3}
\frac{d\sigma(k,y)}{d^2kdy} = \frac{\as C_F}{\pi^2}\frac{1}{(2\pi)^2}S_A\int d^2r' \, n_p(\b r,\b r', Y-y)\, |\b I(\b r',\b k,y;Y_0)|^2\,,
\eeq
where now in place of \eq{Igen} we write
\bea\label{Igen10}
&&\b I(\b x'-\b y',\b k,y;Y_0)=\int d^2z_1\, e^{-i\b k\cdot \b z_1} \left(\frac{\b z_1-\b x'}{|\b z_1-\b x'|^2}- \frac{\b z_1-\b y'}{|\b z_1-\b y'|^2}\right)
\left[ 
N^\frac{1}{2}_D(\b x'-\b y',\b b,y;Y_0)\right.\nonumber\\
&&\left. -N^\frac{1}{2}_D(\b x'-\b z_1,\b b,y;Y_0)-N^\frac{1}{2}_D(\b y'-\b z_1,\b b,y;Y_0)+N^\frac{1}{2}_D(\b x'-\b z_1,\b b,y;Y_0) N^\frac{1}{2}_D(\b y'-\b z_1,\b b,y;Y_0)\right]\,.
\eea

The amplitude $N_D(\b r,\b b,y;Y_0)$ equals to the cross section of single diffractive dissociation of a dipole of transverse size $\b r$, rapidity $y$ and impact parameter $\b b$ on a target nucleus. It satisfies the Kovchegov--Levin  evolution equation \cite{Kovchegov:1999ji}
\bea\label{BL}
&&\frac{\partial N_D(\b x-\b y,\b b, y;Y_0)}{\partial y}\nonumber\\
&&=\frac{2\as C_F}{\pi^2}\int d^2z \, \left[ \frac{(\b x-\b y)^2}{(\b x-\b z)^2(\b y-\b z)^2}-2\pi \delta(\b y-\b z)
 \ln(|\b x-\b y|\Lambda)\right] N_D(\b x-\b z,\b b,y;Y_0)
\nonumber\\
&+&
\frac{\as C_F}{\pi^2}\int d^2z \frac{(\b x-\b y)^2}{(\b x-\b z)^2(\b y-\b z)^2}
\left[N_D(\b x-\b z,\b b,y;Y_0)N_D(\b y-\b z,\b b,y;Y_0)\right.
\nonumber\\
&-& \left. 4 N_D(\b x-\b z,\b b,y;Y_0)N(\b y-\b z,\b b,y)+
2N(\b x-\b z,\b b,y)N(\b y-\b z,\b b,y)\right]\,,
\eea
with the initial condition
\beq\label{indif}
N_D(\b r,\b b,y=Y_0;Y_0)=N^2(\b r, \b b , Y_0)\,.
\eeq
Diffractive gluon production of the kind shown in \fig{fig:diffract2} requires a dedicated study and will certainly lead to a number of interesting observations. We are going to perform such analysis in future publications. In this paper we concentrate on the case $y=Y_0$.

%%%%%%%%%%%%%%%%%%%%%%%%%%%%%%%%%%%%%%%%
\section{Dipole evolution in onium and nucleus}\label{sec:dipole-ev}

\subsection{Dipole evolution in onium}

Dipole evolution in onium is encoded in the function $n_1(\b r, \b r',\b b, y)$
and is determined by solving equation \eq{BFKL} with the initial condition \eq{in.cond}. In the case of a cylindrical profile we use function $n_p(\b r, \b r', y)$ instead. The general solution of the BFKL equation for $n_p(\b r, \b r', y)$ reads
\beq
n_p(\b r,\b r',y)=\int_{-\infty}^\infty d\nu \, e^{2\bas\chi(\nu)y}\, (r/r')^{1+2i\nu}\, C_\nu^p\,,
\eeq
where $\bas = \as N_c/\pi$ and the leading BFKL eigenvalue
\beq\label{chi}
\chi(\nu)=\psi(1)-\frac{1}{2}\psi(\frac{1}{2}-i\nu)-\frac{1}{2}\psi(\frac{1}{2}+i\nu)\,,
\eeq
with $\psi(\nu)$ being the digamma function
\beq\label{digamma}
\psi(\nu)=\frac{\Gamma'(\nu)}{\Gamma(\nu)}\,.
\eeq
The Mellin image $C_\nu^p$ can be found using the formula
\beq
\delta(\b r-\b r')=\frac{1}{2\pi^2r'^2}\,\int_{-\infty}^\infty d\nu\, (r/r')^{1+2i\nu}\,.
\eeq
The result is
\beq\label{np}
n_p(\b r,\b r',y)=\frac{1}{2\pi^2r'^2}\int_{-\infty}^\infty d\nu \, e^{2\bas\chi(\nu)y}\, (r/r')^{1+2i\nu}\,.
\eeq

Integral over $\nu$ can be done analytically in two important limits. In the leading logarithmic approximation (LLA) we expand the function $\chi(\nu)$ near the minimum at $\nu=0$ as
\beq\label{nuLLA}
\chi(\nu)_{LLA}\approx 2\ln 2-7\zeta(3)\nu^2\,,
\eeq
where $\zeta(z)$ is the Riemann zeta function. Substituting \eq{nuLLA} into \eq{np} and integrating around the saddle point
\beq\label{sp2}
\nu_p^{\star} = \frac{i\ln(r/r')}{14\zeta(3)\bas y} \,,
\eeq
we derive
\beq\label{npLLA}
n_p(\b r,\b r',y)_{LLA}\approx \frac{1}{2\pi^2rr'} \sqrt{\frac{\pi}{14\zeta(3)\bas y}}
e^{(\alpha_P-1)y}\, e^{-\frac{\ln^2(r'/r)}{14\zeta(3)\bas y}}\,,\quad \as y\gg \ln^2(r/r')\,,
\eeq
where $\alpha_P-1=4\bas \ln 2$.

Alternatively, we can expand $\chi(\nu)$ near one of its two symmetric poles   at $2i\nu =\pm 1$. This corresponds to the double logarithmic approximation. The choice of a particular pole  depends on the relation between $r$ and $r'$. Expanding near $2i\nu = 1$ we attain 
\beq\label{chiDLA}
\chi(\nu)_{DLA}\approx \frac{1}{1-2i\nu}\,.
\eeq
Plugging this into \eq{np} we have
\beq
n_p(\b r,\b r',y)_{DLA}\approx \frac{1}{2\pi^2r'^2}\int_{-\infty}^\infty d\nu \, e^{\frac{2\bas y}{1-2i\nu}+(1+2i\nu)\ln(r/r')}\,.
\eeq
The saddle point of the expression in the exponent is
\beq\label{spp}
\nu^*_p=\frac{1}{2i}\left( 1-\sqrt{\frac{2\bas y}{\ln(r'/r)}}\right)\,,
\eeq
which is valid only if  $r<r'$. Expanding the argument of the exponential near the saddle point $\nu^*_p$ to the second order and integrating gives the double-logarithmic approximation
\beq\label{npDLA}
n_p(\b r,\b r',y)_{DLA}\approx \frac{r^2}{4\pi^{3/2}r'^4}\frac{(2\bas y)^{1/4}}{\ln^{3/4}(r'/r)}\, e^{2\sqrt{2\bas y\ln(r'/r)}}\,,\quad r<r'\,,\quad \ln(r'/r)\gg \as y\,.
\eeq

To derive an analogous expression at $r>r'$ we expand $\chi(\nu)$ near the symmetric  pole 
\beq\label{chiDLA2}
\chi(\nu)_{DLA}\approx \frac{1}{1+2i\nu}\,.
\eeq
In analogy to \eq{spp} and \eq{npDLA} we derive the saddle point
\beq\label{spp2}
\tilde \nu^*_p=\frac{1}{2i}\left( -1+\sqrt{\frac{2\bas y}{\ln(r/r')}}\right)\,,
\eeq
and the dipole density
\beq\label{npDLA2}
n_p(\b r,\b r',y)_{DLA}\approx \frac{1}{4\pi^{3/2}r'^2}\frac{(2\bas y)^{1/4}}{\ln^{3/4}(r/r')}\, e^{2\sqrt{2\bas y\ln(r/r')}}\,,\quad r>r'\,,\quad \ln(r/r')\gg \as y\,.
\eeq

%%%%%%%%%%%%%%%%%%%%%%%
\subsection{Dipole evolution in a heavy nucleus}

\subsubsection{Leading twist approximation}

Consider the forward elastic dipole--nucleus scattering amplitude $N(\b r,\b b, Y)$ satisfying the nonlinear evolution equation \eq{BK}. If the dipole size is much smaller than the saturation scale $Q_s$, then the quantum evolution of the amplitude is governed by the BFKL equation. Therefore in this case, the general solution is
\beq\label{nag}
N(\b r,\b b, y)_{LT}=\int_{-\infty}^\infty d\nu\, e^{2\bas\chi(\nu)y}\, (rQ_{s0})^{1+2i\nu}\,C_\nu^A\,.
\eeq
The Mellin image $C_\nu^A$ of the amplitude $N(\b r,\b b, 0)_{LT}$ is calculated as follows
\bea\label{ca}
C_\nu^A&=&\frac{Q_{s0}}{\pi}\int _0^\infty dr\, (rQ_{s0})^{-2-2i\nu}\, N(\b r, \b b, 0)_{LT}\nonumber\\
&=&  \frac{Q_{s0}}{\pi}\int _0^\infty dr\, (rQ_{s0})^{-2-2i\nu}\, \frac{1}{8}r^2Q_{s0}^2=\frac{1}{8\pi}\frac{1+(1-2i\nu)\ln\frac{Q_{s0}}{\Lambda}}{(1-2i\nu)^2}\,.
\eea
In the last line of \eq{ca} we used the fact that $Q_{s0}$ logarithmically depends on $r$, see \eq{Qsat},\eq{xG}. 
Analogously to the derivation of \eq{npLLA} we obtain in the leading logarithmic approximation
\beq\label{naLLA}
N(\b r,\b b, y)_{LLA}=\frac{rQ_{s0}}{8\pi}\sqrt{\frac{\pi}{14\zeta(3)\bas y}}\ln\left(\frac{Q_{s0}}{\Lambda}\right) \, e^{(\alpha_P-1)y}\, e^{-\frac{\ln^2(rQ_{s0})}{14\zeta(3)\bas y}}\,,\quad \as y\gg \ln^2\left(\frac{1}{rQ_{s0}}\right)\,,
\eeq
where the saddle point is
\beq\label{sp2a}
\nu_A^\star = \frac{i\ln(rQ_{s0})}{14\zeta(3)\bas y}\,.
\eeq

In the double logarithmic approximation \eq{chiDLA} the saddle point for the case $r<1/Q_{s0}$ is
\beq\label{spa}
\nu^*_A=\frac{1}{2i}\left( 1-\sqrt{\frac{2\bas y}{\ln \frac{1}{rQ_{s0}}}}\right)\,.
\eeq
Repeating the by now familiar procedure we write  
\bea\label{naDLA}
N(\b r,\b b, y)_{DLA}=\frac{\sqrt{\pi}}{16\pi}\frac{\ln^{1/4}\left(\frac{1}{rQ_{s0}}\right)}{(2\bas y)^{3/4}}\,r^2Q_{s0}^2\,\left( 1+  \sqrt{\frac{2\as y}{\ln\frac{1}{rQ_{s0}}}} \, \ln\frac{Q_{s0}}{\Lambda}  \right) e^{2\sqrt{2\bas y\ln\frac{1}{rQ_{s0}}}}\,,&&\nonumber\\
 r<1/Q_{s0}\,,\quad \ln\frac{1}{rQ_{s0}}\gg \as y\,.\qquad &&
\eea

Next, we consider  the case $r>1/Q_{s}$.

%%%%
\subsubsection{Deep saturation region}\label{sec:Nsat}

Solution to the BK equation \eq{BK} deeply in the saturation regime was found in \cite{Levin:1999mw,Levin:2000mv,Levin:2001cv}. With the logarithmic accuracy the dominant dipole splitting corresponds to the configuration in which the size of one of the daughter dipoles ($\sim 1/Q_s$) is much smaller than the other (see \sec{sec:lowx}). Denote again $\b r=\b x-\b y$ and $\b w=\b z-\b y$. In the saturation region we have either $w\ll r \approx |\b w-\b r|$ or the symmetric configuration $|\b w-\b r|\ll r \approx w$. Both give equal contribution to the integral over $\b w$. Restricting ourself to the case $w\ll r$ and doubling the integral we write the BK equation as follows:
\beq\label{xyz}
\frac{\partial N(\b r,\b b, y)}{\partial y}\approx \frac{\as C_F}{\pi}\,2\,\int_{1/Q_s^2}^{r^2}\frac{dw^2}{w^2}\,[N(\b w,\b b, y)-N(\b w,\b b, y)N(\b r,\b b, y)]\,.
\eeq
Now, for the reason that in the saturation region, the amplitude $N(\b r,\b b, y)$ is close to unity we render  \eq{xyz} as 
\beq\label{bksat}
-\frac{\partial\{1- N(\b r,\b b, y)\}}{\partial y}\approx\frac{\as C_F}{\pi}\,2\,\int_{1/Q_s^2}^{r^2}\frac{dw^2}{w^2} \{ 1-N(\b r,\b b, y)\}=
\frac{2\,\as C_F}{\pi}\ln(r^2Q_s^2)\, \{1- N(\b r,\b b, y)\} \,.
\eeq
The saturation scale $Q_s(y)$ can be found by equating the argument of the exponent in \eq{naDLA} to a constant which yields \cite{Levin:1999mw,Bartels:1992ix}
\beq\label{b0}
Q_s(y)\approx Q_{s0}e^{2\bas y}\,.
\eeq
Introducing a new scaling variable $\tau= \ln (r^2Q_s^2)$ we solve \eq{bksat} and find the high energy limit of the forward scattering amplitude (in the fixed coupling approximation). It reads  \cite{Levin:1999mw} 
\beq\label{lt}
N(\b r, \b b, y)=1-S_0\, e^{-\tau^2/8}= 1-S_0\, e^{-\frac{1}{8}\ln^2(r^2Q_s^2)}\,,
\eeq
where  we approximated $C_F\approx N_c/2$ in  the large $N_c$ limit. $S_0$ is the integration constant. It determines the value of the amplitude at the critical line $r(y)=1/Q_s(y)$.

%%%%%%%%%%%%%%%%%%%%%%%%%%%%%%%%%
%%%%%%%%%%%%%%%%%%%%%%%%%%%%%%%%%%%
\section{Diffractive cross section in the quasi-classical approximation}
\label{sec:section}

\begin{figure}[ht]
\begin{tabular}{ccc}
      \includegraphics[width=6cm]{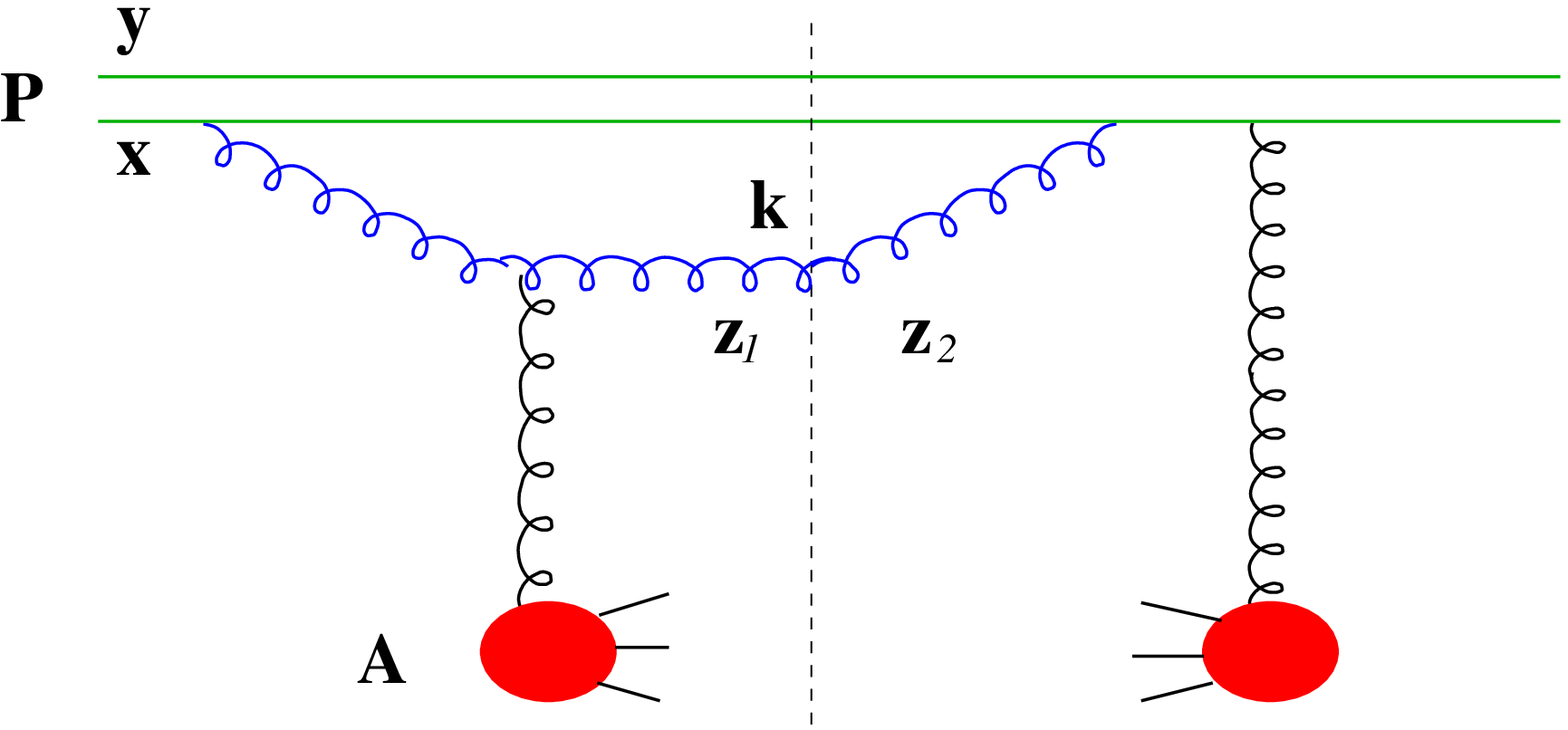} &
     \mbox{$\left. \qquad \right.$ }&
      \includegraphics[width=6cm]{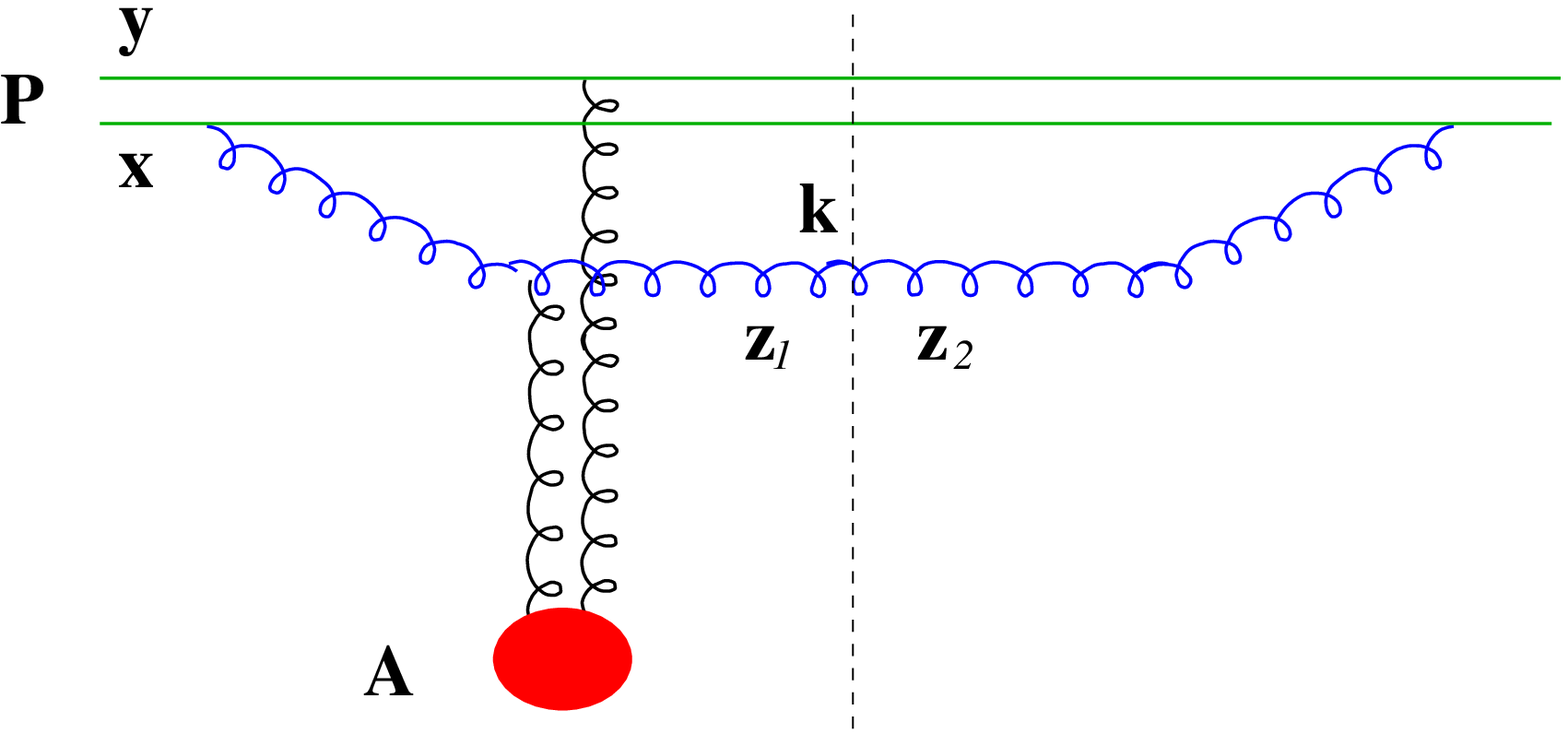} \\
       (a) && (b)
      \end{tabular}
\caption{An example of diagrams contributing to (a) inclusive gluon production and (b) diffractive gluon production.   }
\label{fig:sample}
\end{figure}

Careful inspection of \eq{xsectQ} reveals that the cross section vanishes when size of the onium $\b r=\b x-\b y$ is much larger than the characteristic scale $1/Q_{s0}$. This is in a sharp contrast with the inclusive gluon production case \cite{Kovchegov:1998bi} where the cross section stays finite at $r\to \infty$. To understand the reason for such different behavior, consider a sample diagrams contributing to each of the processes shown in \fig{fig:sample}.
%%%%
%%%%%
In diagram (a), corresponding to the inclusive case, the propagator of the $q\bar qg$ system in the nucleus is proportional to $e^{-\frac{1}{4}(\b x-\b z_1)^2 Q_{s0}^2}$ while the gluon emission amplitude is proportional to $\frac{g\,(\b x-\b z_1)}{|\b x-\b z_1|^2}$. Both do not involve the $\b y$ coordinate at all and are finite at $\b y\to \infty$. On the contrary, in diagram (b), corresponding to the diffractive case, the propagator involves both $\b x$ and $\b y $ coordinates, see \eq{P}, no factorization of $\b y$ dependence similar to the inclusive case happens. All other diagrams contributing to these two processes can be analyzed in the same way. Let us look at the  color structure of the $q\bar q g$ system in the two cases. By Pomeranchuk theorem, the exchanged (Coulomb) gluons are in the color singlet state. Therefore,  we notice that in  diagram (a) the $q\bar q g$ system is in the color octet state and its propagator through the nucleus equals the propagator  of a gluon dipole, whereas in diagram (b) the $q\bar q g$ system is always in the color singlet state corresponding to the quark dipole. This feature can be seen also in expressions for  the propagator, one involving the gluon saturation scale $Q_{s0}^2$ (inclusive case), another involving the quark saturation scale $\frac{C_F}{N_c}Q_{s0}^2\approx \frac{1}{2}Q_{s0}^2$ (diffractive case). 

Now we would like to determine how does the diffractive cross section behave in the quasi-classical approximation. Consider the following integral appearing in the r.h.s.\ of \eq{sectot} and \eq{main-tot}:
\bea\label{J}
J(\b r,y)&=&\frac{1}{\pi}\int d^2w\,\frac{ \b r^2}{(\b w-\b r)^2\b w^2}\,
\left[ N(\b r,\b b,y) \right.   \nonumber\\
&&\left. -N(\b w-\b r,\b b,y)-N(\b w,\b b,y)+N(\b w-\b r,\b b,y)N(\b w,\b b,y)\right]^2\,.
\eea
Let us analyze its behavior in the quasi-classical case ($y=0$) for small $r<1/Q_{s0}$ and large $r>1/Q_{s0}$ onium sizes.

%%%%%%%%
\subsection{Dilute regime $r<1/Q_{s0}$}\label{sec:lin-qc}

In the case of small onium we  divide the entire integral over $\b w$ into three terms as follows:
\bea\label{va1}
J(\b r,0)&\approx & \int_0^r\frac{dw^2}{w^2}\,\left[1-N(\b r, \b b,0)\right]^2\, N^2(\b w, \b b, 0)\nonumber\\
&&
\,+\, r^2\int_r^{1/Q_{s0}}\frac{dw^2}{w^4}\,\left[ N(\b r, \b b,0)-
2N(\b w, \b b, 0)+N^2(\b w, \b b, 0)\right]^2\nonumber\\
&& +\,r^2\int_{1/Q_{s0}}^\infty\frac{dw^2}{w^4}\,\left[ N(\b r, \b b,0)-
2N(\b w, \b b, 0)+N^2(\b w, \b b, 0)\right]^2\,,
\eea
where \eq{va1} holds in logarithmic approximation.
We can estimate each term utilizing the fact that according to  \eq{NQ}
\beq\label{va2}
N(\b r, \b b, 0)\approx 
\left\{ \begin{array}{cc}
1\,, \quad & r\gg 1/Q_{s0}\,,\\
\frac{1}{8}\,r^2Q_{s0}^2\,, \quad & r\ll 1/Q_{s0}\,.
\end{array}
\right.
\eeq
Moreover,  in the last two terms in \eq{va2} $N(\b r, \b b,0)$ can be neglected since in most of the integration regions  $w\gg r$. Indeed, in the second term in the r.h.s.\ of \eq{va1} it can be seen once we neglect $N^2(\b w, \b b,0)$ and expand $N(\b w, \b b,0)$ at small $w$. In the third term in the r.h.s.\ of \eq{va1} we have $2N(\b w, \b b,0)-
N^2(\b w, \b b,0)\approx 1$ while $N(\b r, \b b,0)\ll 1$.
Accordingly, expanding the integrands using \eq{va2} we find 
 that the first term in the r.h.s.\ of \eq{va1} is of the order $\mathcal{O}(r^8 Q_{s0}^8)$, whereas the second and the third ones are of the order $\mathcal{O}(r^2 Q_{s0}^2)$. Therefore, the last two terms in \eq{va1} dominate in the regime $rQ_{s0}\ll 1$. It is customary to denote 
\beq\label{ng}
N_G(\b r, \b b,y) = 2N(\b r, \b b,y)-N^2(\b r, \b b,y)\,,
\eeq
 which has the meaning of the \emph{gluon dipole} forward elastic scattering amplitude. In terms of this quantity the function $J(\b r, 0)$ reads
\beq\label{va3}
J(\b r,0)\approx r^2 \int_0^\infty \frac{dw^2}{w^4}\, N_G^2(\b w, \b b , 0)\,,\quad r\ll 1/Q_{s0}\,,
\eeq
where the lower limit of integration ($r$) has been set to zero with logarithmic accuracy (note that the second integral in the r.h.s. of \eq{va1} is dominated by dipoles of size $w\sim 1/Q_{s0}\gg r$). 

It is useful to notice, that \eq{va3} holds also in the case the low-$x$ evolution is taken into account. In the quasi-classical approximation the integral \eq{va3} can be done if we treat $Q_{s0}$ as a $w$-independent constant neglecting its logarithmic variation. In that case, substituting \eq{NQ} we derive 
\beq\label{va4}
J(\b r,0)\approx \frac{1}{4}\ln 2 \, r^2 Q_{s0}^2\,, \quad r\ll 1/Q_{s0}\,,
\eeq
The cross section is obtained using \eq{sectot} and \eq{J}. We have
\beq\label{va5}
\frac{d\sigma}{dy}=\frac{\as C_F}{\pi }\,S_A\, J(r,0)= \frac{\as C_F\ln 2}{4\pi }\,S_A\,
r^2Q_{s0}^2\,, \quad r\ll 1/Q_{s0}\,.
\eeq

%%%%%%%%%%%%%%%%%%%%%%%%%%%%%%
\subsection{Dense regime $r>1/Q_{s0}$}\label{sec:quasi-dense}

As in the previous case we divide the integral into three parts
\bea\label{vb1}
J(\b r,0)&\approx & \int_0^{1/Q_{s0}}\frac{dw^2}{w^2}\,\left[1-N(\b r, \b b,0)\right]^2\, N^2(\b w, \b b, 0)\nonumber\\
&&
\int_{1/Q_{s0}}^r\frac{dw^2}{w^2}\,\left[1-N(\b r, \b b,0)\right]^2\, N^2(\b w, \b b, 0)\nonumber\\
&& +\,r^2\int_r^\infty\frac{dw^2}{w^4}\,\left[ N(\b r, \b b,0)-
2N(\b w, \b b, 0)+N^2(\b w, \b b, 0)\right]^2\,.
\eea
Utilizing \eq{va2} we simplify \eq{vb1} in the logarithmic approximation as follows
\bea
J(\b r,0)&\approx& \left[ N(\b r, \b b,0)-
1\right]^2 \, \left(
 \int_0^{1/Q_{s0}}\frac{dw^2}{w^2}\,\frac{1}{64}w^4Q_{s0}^4 +
\int_{1/Q_{s0}}^r\frac{dw^2}{w^2}\,  +\,r^2\int_r^\infty\frac{dw^2}{w^4}\right)\,.
\label{vb2}\\
&\approx & \left[ N(\b r, \b b,0)-
1\right]^2 \ln(r^2Q_{s0}^2)\label{vb3}\,,\quad r\gg 1/Q_{s0}\,.
\eea
Going from \eq{vb2} to \eq{vb3}  we kept only the second term in the brackets in \eq{vb2} as it is logarithmically enhanced.  Formula \eq{vb3} is valid in the case of low-$x$ evolution as well. Substituting \eq{NQ} into \eq{vb3} yields
\beq\label{vb4}
J(r)=\ln(r^2Q_{s0}^2)\, e^{-\frac{1}{4}r^2Q_{s0}^2}\,,\quad r\gg 1/Q_{s0}\,.
\eeq
Finally, the cross section follows from \eq{sectot}, \eq{J} and \eq{vb4} as
\beq\label{vb5}
\frac{d\sigma}{dy}=\frac{\as C_F}{\pi}\,S_A\, \ln(r^2Q_{s0}^2)\, e^{-\frac{1}{4}r^2Q_{s0}^2}\,,\quad r\gg 1/Q_{s0}\,.
\eeq
The striking feature of this formula is strong exponential suppression of diffractive gluon production for large onium.  We will see in the next section that this result completely changes when the quantum evolution in the onium becomes an important effect.

%%%%%%%%%%%%%%%%%%%%%%%%%%%%%%%%%
%%%%%%%%%%%%%%%%%%%%%%%%%%%%%%%%%%%

\section{Diffractive cross section including low-$x$ evolution}
\label{sec:lowx}

Using \eq{main-tot} and \eq{J} we write
\beq\label{via1}
\frac{d\sigma}{dy}=\frac{\as C_F}{\pi}\,S_A\int d^2r'\, n_p(\b r, \b r', Y-y)\, J(\b r',y)\,.
\eeq

\subsection{Dilute regime $r<1/Q_s(y)$}\label{sec:le1}

As in the quasi-classical case, first  we are going to find the kinematic region which gives the largest (logarithmic) contribution to the  integral. We have
\bea\label{via2}
\frac{d\sigma}{dy}&=& \frac{\as C_F}{\pi}\,S_A\,2\pi \left[ \int_0^r dr'r' n_p(\b r, \b r', Y-y)\, J(\b r',y)\right.\nonumber\\
&& \left. +\int_r^{1/Q_s} dr'r' n_p(\b r, \b r', Y-y)\, J(\b r',y)+
\int_{1/Q_s}^\infty dr'r' n_p(\b r, \b r', Y-y)\, J(\b r',y)\right]
\eea
As has been noted in the previous sections, equations \eq{va3} and \eq{vb3} hold also in the evolution case, provided the $y$-dependence is explicitly indicated in the arguments of $J(\b r,y)$ and $N(\b r, \b b, y)$. Generalization of \eq{va4} reads
\beq\label{via3}
J(\b r', y)\approx C_0\, r'^2\, Q_s^2(y)\,, \quad r'\ll 1/Q_s(y)\,,
\eeq
where $C_0$ is a constant which depends  on a particular  functional form of $N_G(\b r, \b b , y)$ and can be found numerically from \eq{BK}.  Using \eq{lt} in \eq{vb3} gives another limit of function $J(\b r', y)$:
\beq\label{via4}
J(\b r', y)\approx S_0^2\, e^{-\ln^2(r'Q_s)}\, \ln (r'^2Q_s^2)\,, \quad r'\gg 1/Q_s(y)\,.
\eeq

Accordingly, using \eq{npDLA} or \eq{npDLA2}  depending on the relation between $r$ and $r'$, i.\ e.\ $n_p\sim r^2/r'^4$ if $r<r'$ or $n_p\sim 1/r'^2$ if $r>r'$, as well as \eq{via3} and \eq{via4} we
estimate that the second integral in \eq{via2} is enhanced by  $\ln \frac{1}{rQ_s}$ with respect to the first one, whereas the third integral is vanishingly small. Thus, 
\bea\label{via5}
\frac{d\sigma}{dy}&\approx& \frac{\as C_F}{\pi}\,S_A\,2\pi \int_r^{1/Q_s} dr'r' n_p(\b r, \b r', Y-y)\, J(\b r',y)\nonumber\\
&=& \frac{2\, C_0\,\as C_F S_A}{4\pi^{3/2}} \, r^2 \, Q_s^2\,\int _r^{1/Q_s} \frac{dr'}{r'}\,\frac{(2\bas (Y-y))^{1/4}}{\ln^{3/4}\frac{r'}{r}}\,e^{2\sqrt{2\bas (Y-y)\ln \frac{r'}{r}}}\,.
\eea
Changing to a new integration variable $\eta$ defined as  $\eta^2=\ln\frac{r'}{r}$  the integral in \eq{via5} can be taken explicitly in terms of the imaginary error function. In the double-logarithmic approximation the result reads
\beq\label{via6}
\frac{d\sigma}{dy}=\frac{C_0\,\as C_F S_A}{4\pi^{3/2}} \, 
\frac{r^2 \, Q_s^2}{(2\bas (Y-y)\, \ln\frac{1}{rQ_s})^{1/4}}\, 
e^{2\sqrt{2\bas (Y-y)\ln \frac{1}{rQ_s}}}\,, \quad r\ll 1/Q_s(y)\,.
\eeq
Both the quasi-classical result \eq{va5} and its quantum counterpart \eq{via6} show that the cross section is proportional to $r^2$ as required by the color transparency. 

%%%%%%%%%%%%%%%%%%%%%%%%%%%%%%
\subsection{Dense regime $r>1/Q_{s}$}

Analogously to \eq{via2} we get
\bea\label{vib1}
\frac{d\sigma}{dy}&=& \frac{\as C_F}{\pi}\,S_A\,2\pi \left[ \int_0^{1/Q_s} dr'r' n_p(\b r, \b r', Y-y)\, J(\b r',y)\right.\nonumber\\
&& \left. +\int^r_{1/Q_s} dr'r' n_p(\b r, \b r', Y-y)\, J(\b r',y)+
\int_r^\infty dr'r' n_p(\b r, \b r', Y-y)\, J(\b r',y)\right]\,.
\eea
The logarithmically enhanced contribution arises from the second integral which -- upon substitution of \eq{via4} and \eq{npDLA2} -- becomes
\beq\label{vib2}
\frac{d\sigma}{dy}= \frac{\as C_F S_A}{2\pi^{3/2}}\, S_0^2\,(2\bas (Y-y))^{1/4}\,
\int_{1/Q_s}^r\frac{dr'}{r'}\frac{\ln(r'^2Q_s^2)}{\ln^{3/4}\frac{r}{r'} }\, e^{-\ln(r'Q_s)}\,e^{2\sqrt{2\bas (Y-y)\ln \frac{r}{r'}}}\,.
\eeq
Note that in the relevant kinematic region $1/Q_s\ll r'\ll r$ we can approximate  $
\ln\frac{r}{r'}=\ln(rQ_s)+\ln\frac{1}{r'Q_s}\approx \ln(rQ_s)$.
The integral over $r'$ then becomes trivial yielding the final result
\beq\label{vib3}
\frac{d\sigma}{dy}= \frac{\as C_F S_A}{2\pi^{3/2}}\, S_0^2\,\frac{(2\bas (Y-y))^{1/4}}{\ln^{3/4}(rQ_s)}\,e^{2\sqrt{2\bas (Y-y)\ln (rQ_s)}}\,, \quad r\gg 1/Q_s(y)\,.
\eeq
We observe  that the cross section given by \eq{vib3} is an \emph{increasing }function of the rapidity interval $Y-y$ between the onium and the nucleus. Together with  the quasi-classical expression \eq{vb5} it implies that the total  cross section for the diffractive gluon production in a scattering of a large onium off the heavy nucleus is non-vanishing only if the low-$x$ evolution in onuim is  an important effect. This can be seen directly in \fig{fig:diffract1}: the gluon multiplicity arises from the cut Pomeron attached to the onium. 
This observation has important phenomenological consequences as we discuss in the next section.

%%%%%%%%%%%%%%%%%%%%%%%%%%%%%%%%%
\section{Discussion and summary} \label{sec:summary}

In this paper we discussed the coherent diffractive gluon production in high energy onium-nucleus collisions. The gluon multiplicity in the case of onium of small  size $r<1/Q_s$ is given by \eq{va5} and \eq{via6} and can be summarized as follows
\beq\label{s1}
\frac{d N_D(y)}{dy}\propto r^2\, Q_s^2(y)\, xG\left( \exp(y-Y),Q_s^2\right)\,, \quad r\ll Q_s\,,
\eeq
where $xG(x,Q^2)$ is a gluon distribution function at  momentum scale $Q^2$. Gluon multiplicity vanishes in the limit $r\to 0$ as is required by the color transparency.

In the other limit of large onium, the gluon production cross section vanishes in the quasi-classical approximation as implied by \eq{vb5}.  At $\bas (Y-y)\gtrsim 1$ the evolution effects in onium play increasingly important role. It is the cut  Pomeron, connecting the onium and the dipole ($\b r'$) emitting the triggered gluon, which contributes to the fast increase in gluon multiplicity  as interval $Y-y$ increases. One way to see it is to recall that during the linear evolution dipoles of various sizes are produced from the parent onium of size $r$. We explained in \eq{vib1} and \eq{vib2}  that the main contribution to the multiplicity stems from the dipoles of size $r'\sim 1/Q_s$ no matter how big is the initial dipole $r$. 
The resulting expression \eq{vib3} has the following behavior
\beq\label{s2}
\frac{d N_D(y)}{dy}\propto  xG\left( \exp(y-Y),Q_s^2\right)\,, \quad r\gg Q_s\,, \,\,\, (Y-y)\gtrsim 1/\as\,. 
\eeq

\begin{figure}[ht]
  \includegraphics[width=8cm]{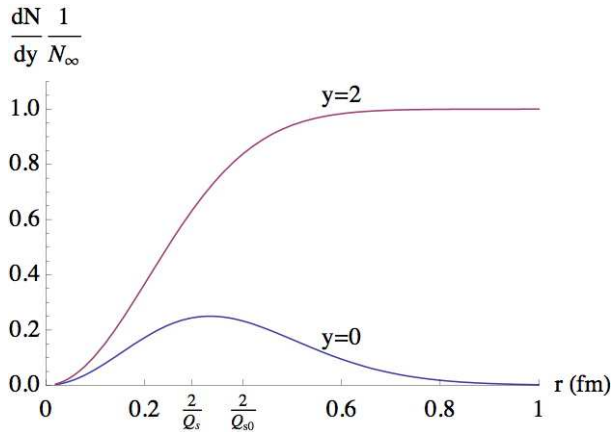}
\caption{Relative multiplicity of diffractive gluons as a function of the onium size $r$ in the quasi-classical approximation (labeled $y=0$) and at very low $x$ ($y=2$). The rapidity values correspond to those at RHIC as explained in the text.    We chose $Q_{s0}=1$~GeV and $Q_{s}=1.35$~GeV. $N_\infty$ is a normalization constant.}
\label{summ}
\end{figure}

Dependence of the diffractive gluon multiplicity on the onium size is summarized in the \fig{summ}.
%%%%
%%%%%

To the extent that the large onium can serve as a model for proton, \eq{vb5} and \eq{vib3} describe the diffractive gluon production in proton-nucleus collisions. 
As such it has a direct implications to the RHIC and LHC phenomenology. 
Phenomenological studied show that the gluon saturation at RHIC starts to impact the gluon and valence quark spectra at rapidities $\eta\simeq 1$ (and larger). In our notations it corresponds to the rapidity interval $y\simeq 6$ between the gluon and the heavy nucleus and $Y-y\simeq 4$ between the proton and the gluon. This corresponds to $x_p\simeq e^{-4}\approx 0.02$ which is perhaps insufficient to have a sizable low-$x$ effect in proton implying a very low multiplicity of diffractive gluon production. On the other hand, exploring the backward rapidity region $\eta<0$ will not allow to probe the gluon saturation in the nucleus. Therefore, if the typical inter-quark  distance in proton  is larger than $\sim 1/Q_s\simeq 0.2$~fm, then we do not expect a significant multiplicity of   gluons in \emph{coherent} diffraction of a proton on nucleus at RHIC, see \fig{summ}. \footnote{Clearly, this conclusion does not hold if proton's valence quarks form a  diquark-quark configuration where the diquark has small size. 
From this perspective the diffractive gluon production at RHIC is a sensitive probe of the valence quark configuration in proton. }

The situation radically  changes at LHC where an  additional rapidity window $\Delta \eta \simeq 6$ opens up. From the point of view of gluon saturation, the mid-rapidity in pA at LHC is expected to be similar to the rapidity $\eta=3$ at RHIC \cite{Kharzeev:2003wz,Tuchin:2007pf}. At the same time, at the LHC midrapidity, $x_p\simeq e^{-7}= 0.001$ which is certainly sufficient for the low-$x$ evolution to take place in proton. Therefore, we expect that measurements of the diffractive gluon production in pA collisions at LHC will be a sensitive probe of the low-$x$ dynamics. At EIC the typical dipole size $r$ is determined by the photon virtuality $Q$ as $r\sim 1/Q$ which makes it possible the detailed study gluon saturation using the diffractive gluon production in different kinematical regions (see e.g.\ \cite{Gotsman:2000zk,Gotsman:2000fy}).

An interesting extension of our work is a case of diffractive production with small rapidity gap $Y_0<y$, which will be relevant at LHC.   Of special interest is dependence of the differential cross section for diffractive gluon production on transverse momentum of produced gluon. We are addressing  this and other issues  in the forthcoming publication.

%%%%%%%%%%%%%%%%%%%%%%%%%%%%%%%%
\acknowledgments
K.T.\ is grateful to  Yuri Kovchegov, Genya Levin  and Jianwei Qiu for very informative and helpful discussions.  
The work of K.T. was supported in part by the U.S. Department of Energy under Grant No.\ DE-FG02-87ER40371. He would like to
thank RIKEN, BNL, and the U.S. Department of Energy (Contract No.\ DE-AC02-98CH10886) for providing facilities essential
for the completion of this work.

%%%%%%%%%%%%%%%%%%%%%%%%%%%%%%%%%%%%%

\end{document}